\documentclass[reprint,prb,superscriptaddress,nobibnotes]{revtex4-1}
\usepackage{amssymb,amsmath}
\usepackage{graphicx}
\usepackage{dcolumn}
\usepackage{multirow}
\usepackage{color}

\definecolor{purple}{rgb}{0.61,0.19,1.00}

\begin{document}

\title{\textbf{\fontfamily{phv}\selectfont Compressed optimization of device architectures}}
\author{Adam Frees\footnote{test}}
\affiliation{These authors contributed equally to this work.}
\affiliation{Department of Physics, University of Wisconsin-Madison, Madison, WI 53706}
\affiliation{Center for Computing Research, Sandia National Laboratories, Albuquerque, NM 87123}
\author{John King Gamble}
\email[Correspondence should be addressed to: ]{jkgambl@sandia.gov}
\affiliation{These authors contributed equally to this work.}
\affiliation{Center for Computing Research, Sandia National Laboratories, Albuquerque, NM 87123}
\author{Daniel R. Ward}
\affiliation{Department of Physics, University of Wisconsin-Madison, Madison, WI 53706}
\affiliation{Center for Computing Research, Sandia National Laboratories, Albuquerque, NM 87123}
\author{Robin Blume-Kohout}
\affiliation{Center for Computing Research, Sandia National Laboratories, Albuquerque, NM 87123}
\author{M. A. Eriksson}
\author{Mark Friesen}
\author{S. N. Coppersmith}
\affiliation{Department of Physics, University of Wisconsin-Madison, Madison, WI 53706}

\maketitle
\textbf{
Recent advances in nanotechnology have enabled researchers to control individual quantum mechanical objects with unprecedented accuracy, opening the door for both quantum\cite{Petta:2005p2180,Koppens:2005p717,PhysRevLett.108.046808,Maune:2012p344,Shulman:2012p202,Pla:2012p541,Kim:2014p70,Wu:2014p11938} and extreme-scale conventional computing applications\cite{Fuechsle:2012p242}.
As these devices become  
larger and more complex, 
the ability to design them such that they can be simply controlled
becomes a daunting and computationally infeasible task\cite{687879}.
Here, motivated by ideas from compressed sensing\cite{Candes:2006p1207,Donoho:2006p1289}, we introduce 
a protocol for the Compressed Optimization of Device Architectures (CODA). 
It leads naturally to 
a metric for benchmarking device performance and optimizing device designs, 
and provides a scheme for automating the control of gate operations and reducing their complexity.
Because CODA
is computationally efficient, it
is readily extensible to large systems.
We demonstrate the CODA benchmarking and optimization protocols through simulations of up to eight quantum dots in devices that are currently being developed experimentally 
for quantum computation.
}

Nanoscale devices are challenging to control in part because their size makes them susceptible to even the smallest material defects.
Quantum devices are especially challenging because their energy spectra and tunnel couplings both require fine tuning\cite{Hanson:2007p1217}.
Here we focus on quantum bits (qubits) formed of electron spins in
electrostatically-gated quantum dots\cite{Loss:1998p120} and donors\cite{Kane:1998p133}. 
Such devices are extensible to
large arrays by leveraging the mature semiconductor processing industry.
Working in a variety of systems, researchers have already demonstrated complete control and excellent decoherence properties of devices
with up to four quantum dots\cite{Petta:2005p2180,Koppens:2005p717,PhysRevLett.108.046808,Maune:2012p344,Shulman:2012p202,Medford:2013p654,Medford:2013p050501,Wu:2014p11938,Kim:2014p70,Kawakami:2014p666,Veldhorst:preprint2013,Enge1500214,KimNatNano15,Kim:2015aa, PhysRevLett.116.046802, 2016arXiv160207833T,PhysRevLett.116.116801}, as well as impurities coupled to electronic reservoirs\cite{Pla:2012p541,Fuechsle:2012p242,Pla:2013p334,Muhonen:2014p1402.7140,Zwanenburg:2013p961,2015arXiv151201606H,0953-8984-27-15-154205,Lauchte1500022}.

Processing quantum information in semiconductors requires 
fine control of the energies and tunnel rates of the individual electrons.
Voltages are simultaneously tuned on many electrodes to precisely shape the electrostatic potential landscape within a device.
Unfortunately, gate geometries that work well for 1-2 qubits are not necessarily well suited for larger arrays. 
For example, electrodes designed to control a given dot will also affect its neighbors via capacitive crosstalk.
While voltage compensation methods are used to manage crosstalk in smaller devices,
hand tuning becomes impractical for larger arrays.
Recent methods have been proposed to help automate this process, including optimized randomized benchmarking for immediate tune-up (ORBIT)\cite{PhysRevLett.112.240504}, real-time Hamiltonian estimation\cite{Schulman:2014unpub}, and computer-automated device tuning\cite{BaartAPL}.
While promising, these schemes are either software-based or are only directly applicable to specific device designs. In either case, these schemes are constrained by the capabilities of the quantum device hardware.

In this paper, we aim to make the control problem more tractable through hardware optimization.
We show how to   
modify systematically dot properties, such as occupations, energies, and tunnel rates, while changing as few voltages as possible -- a strategy we refer to as \emph{control sparsity}.
The scheme relies on results and methods used for compressed sensing\cite{Candes:2006p1207,Donoho:2006p1289} in the signal processing literature,
in which the decoding of signals can be made much more efficient by exploiting their sparseness.
To demonstrate these concepts, we implement the CODA protocol using realistic simulations of several quantum dot devices.
We show that CODA yields solutions that are simultaneously sparse and spatially localized near the relevant dot -- an extremely desirable property for extensibility.
Moreover, formulating control as an optimization problem
allows us to directly compare the effectiveness of different device architectures, enabling optimization of the device
designs themselves.

\begin{figure*}[tb]
\includegraphics[width= 0.9 \linewidth]{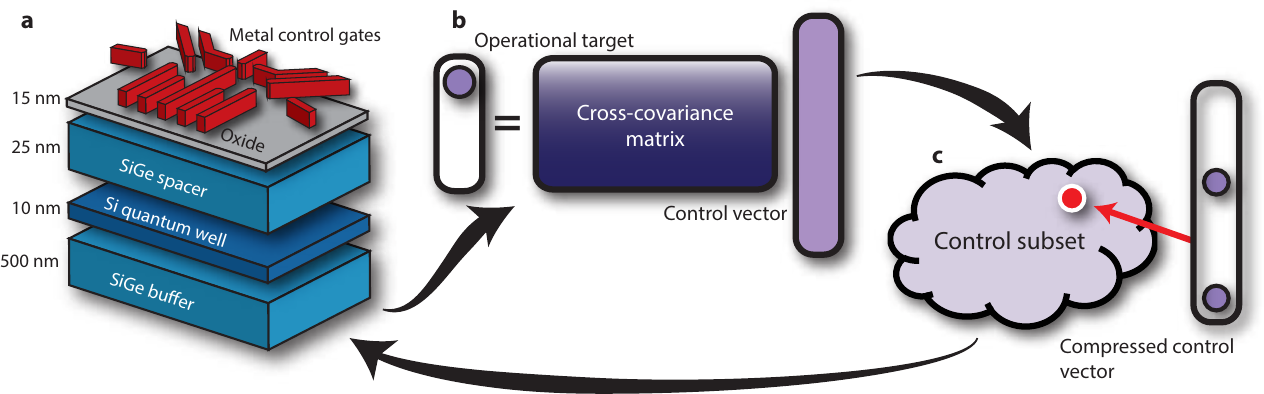}
\caption{\label{fig1} 
Illustration of the CODA procedure for characterizing and optimizing nanoscale device designs. As an example, we consider a Si/SiGe electrostatically defined quantum dot qubit device\cite{Wu:2014p11938,Kawakami:2014p666}. Metal electrodes on the device surface are used to accumulate and control a two-dimensional electron gas (2DEG) in the Si quantum well, forming a double quantum dot.
 \textbf{a}, 
 Device simulations determine the response of the operational targets to the physical controls.
The gate voltages are adjusted to tune the device to an initial working point. Then, the configuration is probed by varying the control gate voltages and noting the response of target variables of interest (e.g. quantum dot occupations and tunnel couplings).
\textbf{b},
A cross-covariance matrix describing the linear response of the system about the working point is determined using the results of (\textbf{a}). 
\textbf{c},
Since there are more controls than targets, the inverse problem yields a subset of valid control vectors for every operational target.
By imposing $L_1$ regularization, we can identify control vectors that are both local and sparse. Global optimization is achieved by using predictions from a single step in the CODA cycle to identify a new working point, and iterating the procedure.}
\end{figure*}

For qubit applications, typical properties we wish to control are the quantum dot occupations and the tunnel rates between the dots, although other properties may also be of interest. 
Such properties are referred to as \emph{operational targets}.
Generally, an operational target could be composed of several physical attributes; we therefore represent it as a vector $\mathbf t$ in a vector space of outputs $\mathcal T$.
The outputs are determined by the voltages applied to the electrodes and by the device structure.
A given set of voltages is referred to as a \emph{control setting}, represented as a vector $\mathbf c$ in a vector space of controls $\mathcal C$.
A physical system is then representable by a function that maps the controls to the targets: $\hat S: \mathcal C \rightarrow \mathcal T$,
shown in Fig.~\ref{fig1}. 
Although this mapping is generally nonlinear, for the moment we limit ourselves to a neighborhood in which the mapping is linear, centered around a working point. 

In an experiment, it is often the case that one particular operational target needs to be pulsed rapidly, while all the others are held fixed.
Since many different $\mathbf c$ can yield a desired $\mathbf t$, we need both a metric that defines the best $\mathbf c$ and an efficient way to find this control.
It is intuitive that solutions are not all equivalently useful: some have extreme voltages that are not experimentally practical,
some involve precisely manipulating all the control voltages of the system, while others involve only modest voltages on a few electrodes.
The latter scenario describes a desirable property often referred to as ``orthogonal" control; \emph{i.e.,} that relatively few electrodes should 
be involved when trying to change an operational target. 
Hence, our goal is to identify solutions in which the resulting change in applied voltages is as simple as possible to implement.

This problem is exactly suited to compressed sensing techniques:
we are given an underdetermined linear mapping $\hat S : \mathcal C~\rightarrow~\mathcal T$, and we wish to find
\begin{equation}
\delta \mathbf c_0 = \underset{\delta \mathbf c} {\text{argmin}} \left|\left| \delta \mathbf c \right|\right|_0,
\mbox{ subject to }\hat S(\mathbf c_\text{op} + \delta \mathbf c) = \mathbf t_\text{op}  + \delta \mathbf t ,\label{eq1}
\end{equation}
where $\mathbf c_\text{op}$  ($\mathbf t_\text{op}$) is the control (target) vector at the working point,
$\delta \mathbf c$ ($\delta \mathbf t$) is the control (target) change from the working point, and 
$|| \cdot ||_0$ is the $L_0$ pseudonorm\cite{Donoho:2006p1289}, which counts the number of non-zero elements in the vector. 
This type of constrained optimization is known as \emph{regularization}. Unfortunately, regularization with respect to the $L_0$ pseudonorm is known to be an NP-hard problem\cite{Natarajan:1995p227}, so finding a solution is not computationally feasible for large systems. Compressed sensing avoids this
difficulty by regularizing the problem using the $L_1$ norm
(the sum of the absolute values of the entries of the vector), which is computationally efficient. Regularizing with the $L_1$ norm tends to achieve sparse solutions, while regularizing with the $L_2$ norm (the square root of the sum of the squares of
the entries of the vector) does not\cite{Candes:2006p1207,Donoho:2006p1289}.

Results from compressed sensing show that the control vectors produced by this method are indeed sparse\cite{Candes:2006p1207,Donoho:2006p1289}; however, the question of locality has not been addressed.
We first give an intuitive argument for why CODA should yield solutions that are local, and then provide a numerical demonstration of the fact.
Consider a classical scenario in which the dot couplings are purely capacitive. 
Since the capacitive coupling between a dot and an electrode decays inversely with separation, electrodes that are far away tend to require large voltage changes to achieve the same response as closer electrodes.
Minimizing the $L_1$ norm of the voltage vector $\delta \bf{c}$ suppresses such changes in favor of small voltage changes.
In other words, the sparse solutions found by CODA are local.

\begin{figure}[tb]
\includegraphics[width= 0.95 \linewidth]{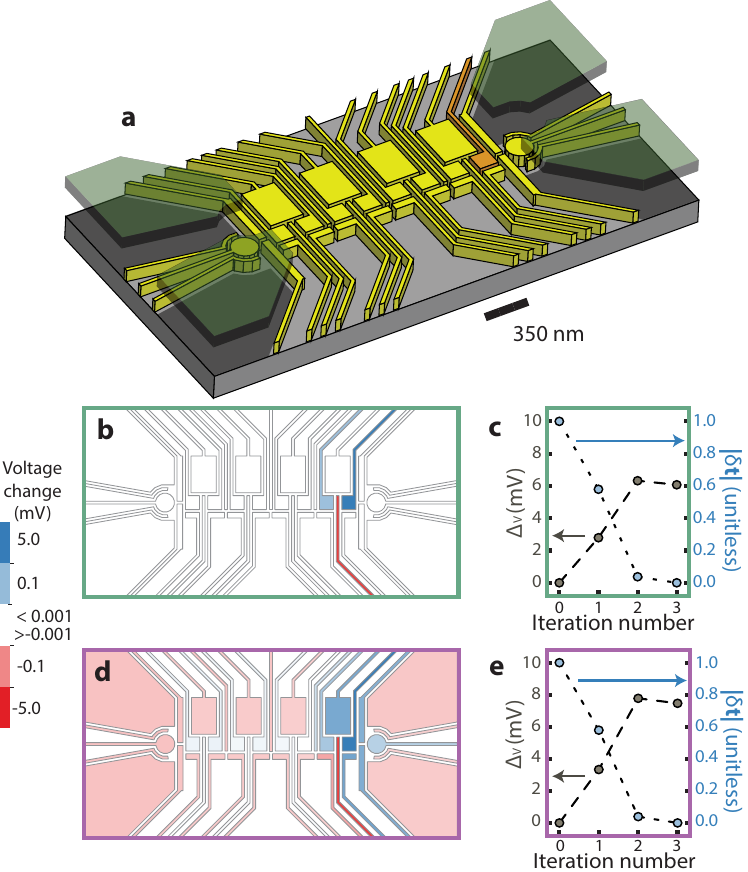}
\caption{\label{octo} 
Locality emerging from the CODA protocol applied to an octuple quantum dot, comprising four double-dot qubits. 
\textbf{a}, Device schematic, with metal gates colored yellow/orange (lower level) and green (upper level). The operational target is to increase the occupation of the right-most quantum dot (underneath the orange electrode) by one electron.\
\textbf{b,d}, Visualization of the voltage variations corresponding to adding one electron to the right-most dot while leaving all other dot occupations and tunnel barriers unchanged, plotted on a logarithmic color scale (electrodes with voltage changes less than 1 $\mu$V are colored white). Solutions are obtained by applying the iterative CODA procedure, minimizing the \textbf{b}, $L_1$ and \textbf{d}, $L_2$ norms of the voltage changes found for each cycle.
\textbf{c,e}, The $L_1$ norm of the voltage changes relative to the original working point ($\Delta_V$) and the associated distance from the operational target ($\left|  \delta \mathbf t \right|$, see Supplemental Information~\cite{suppMat} for details about this metric) across each iteration of the CODA procedure. Solutions are obtained by minimizing the \textbf{c}, $L_1$ and \textbf{e}, $L_2$ norms of the voltage changes found for each cycle.
These results demonstrate that the voltage changes obtained using $L_1$ minimization are local, requiring adjustment of a small number of electrodes, while many electrode voltages are changed when $L_2$ minimization is used.
}
\end{figure}

\begin{figure}[tb]
\includegraphics[width= 0.95 \linewidth]{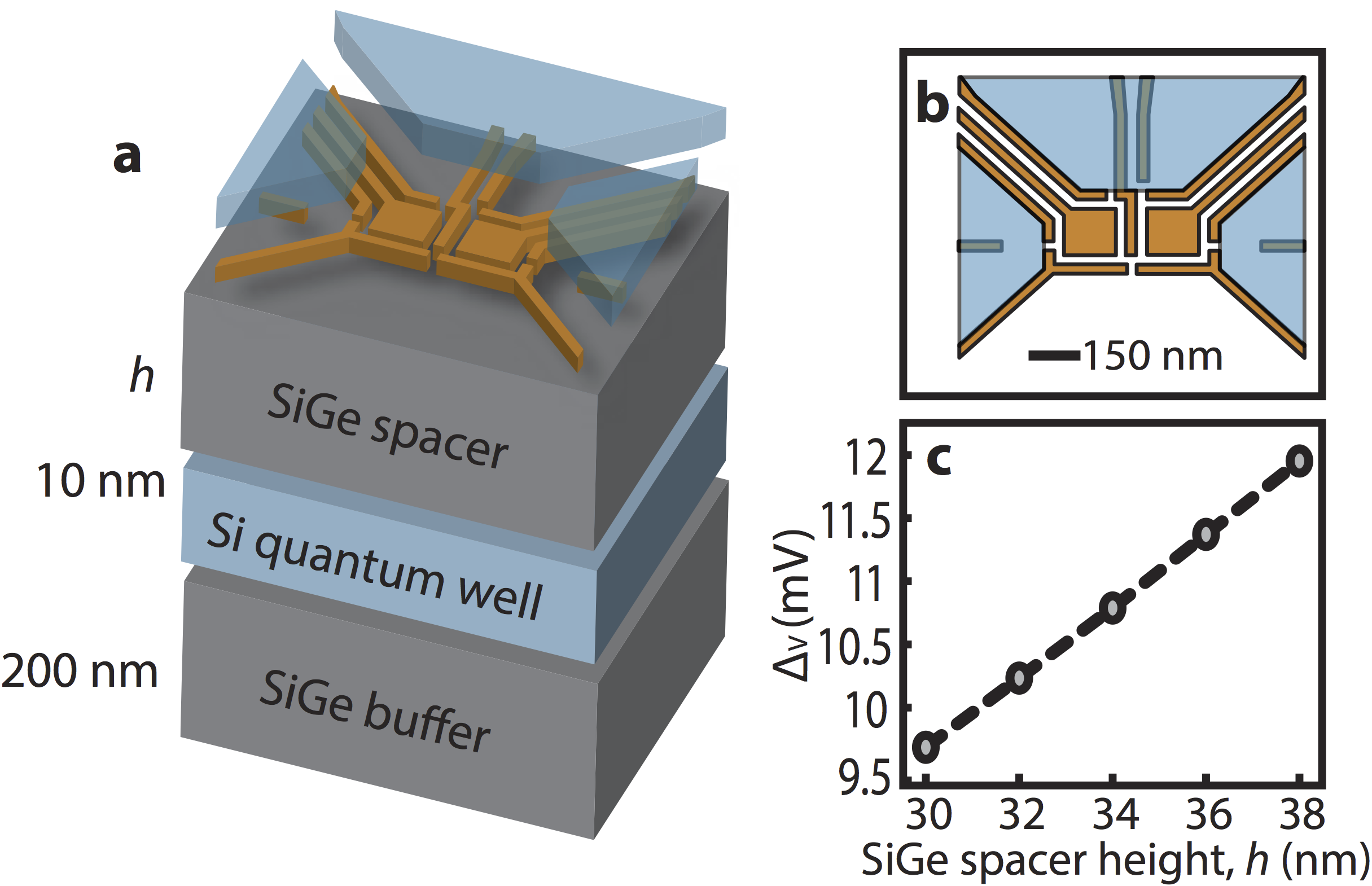}
\caption{\label{varyHeight} 
Improving double quantum dot designs through CODA.
\textbf{a}, A schematic of a series of double-dot devices with different SiGe spacer thicknesses. The heterostructure, which is not shown to scale here, consists of 200 nm of Si$_{0.7}$Ge$_{0.3}$ (the SiGe ``buffer"), 10 nm of Si (the Si quantum well), and a layer of Si$_{0.7}$Ge$_{0.3}$ with variable height $h$ (the SiGe ``spacer").
\textbf{b}, The gate geometry used for the devices. The design consists of an upper layer and lower layer of electrodes, depicted in blue and orange, respectively, separated by 80nm of oxide. The lower layer of electrodes is separated from the SiGe spacer by 10nm of oxide.
\textbf{c}, The CODA procedure is applied to a range of devices with different spacer heights, $h$.  For each simulation we begin with an initial working point corresponding to one electron per dot and an interdot tunnel barrier of 1 meV. We then use CODA to add one electron to the right dot, while keeping the occupation of the left dot and the barrier height fixed. The $L_1$ norm, $\Delta_V$, of the voltage required to achieve this retuning monotonically increases with the height of the SiGe spacer, indicating that thinner spacer layers allow the dots to be controlled more easily.
}
\end{figure}

Figure~\ref{octo} gives a numerical demonstration of the emergence of local solutions when CODA is applied to 
an accumulation-mode 8-dot device in a 
Si/SiGe heterostructure with
four capacitively coupled double quantum dot qubits.
This design represents a natural extension of a recently demonstrated 4-dot device \cite{ward:2016}.
Figure~\ref{octo}a shows the device design.
We model the device using the semiclassical Thomas-Fermi approximation\cite{Stopa:1996p13767} to compute electron densities and potentials, as described in the Methods. The space of operational targets is defined by the eight dot occupations and the four intra-qubit tunnel rates. 
Our working point is chosen so there is 1 electron per dot, and energy barriers with maximum heights of 1 meV, 
corresponding roughly to GHz tunnel rates\cite{2015Nanot..26t5703S} (see Supplemental Information~\cite{suppMat} for details about the simulation parameters). 

We then calculate small changes in the dot occupations and tunnel rates when electrode voltages are varied near the working point, obtaining the cross-covariance matrix that describes the effective linear response of the system.
Once the cross-covariance matrix is determined, we may
specify a desired output.
Here, we focus on the operational target of adding one electron to the right-most quantum dot, while holding all the other tunnel rates and dot occupations constant. 

As this is a linear approximation of a non-linear process, a single cycle of the CODA procedure will not necessarily generate voltages which will yield the desired operational target in the device. Regularization therefore requires a closed-loop iteration, as indicated in Fig.~\ref{fig1}.
To reliably converge, the step size needs to be sufficiently small that the linearized model remains approximately correct. 
In Fig.~\ref{octo}b, we show the voltage changes needed to achieve convergence of the CODA procedure---in this case, after three iterative steps. Note that although voltages are allowed to vary on all the electrodes, only three voltages are found to vary in the $L_1$-regularized solution, corresponding to electrodes that are most local to the target of interest.
Hence, we confirm that the $L_1$ norm is an effective proxy for local control, and that CODA is a practical tool 
for tuning a device,
because it picks out control solutions that are both sparse and local while achieving a specified operational target. For comparison, we also performed an alternative optimization procedure based on $L_2$ regularization, with results shown in Fig.~\ref{octo}d. Note that although the latter procedure achieves the same target objectives, the solution involves voltage changes on almost all of the electrodes, indicating that this solution is neither local nor sparse.

We now show how CODA can be used to optimize device designs, via the objective comparison of different device designs. 
We focus on a single double-dot qubit design, shown in Fig.~\ref{varyHeight}a, which can be replicated to form a multi-qubit device.
Since CODA automatically localizes the control voltages, using it to design a high-quality unit cell leads naturally to multi-qubit devices that are locally controllable.

We again use the semiclassical Thomas-Fermi approximation \cite{Stopa:1996p13767} to model the devices, with working points similar to those used for the 8-dot device.
The heterostructure of the simulated devices is shown in Fig.~\ref{varyHeight}a (see Methods for details).
To demonstrate how CODA can be used to compare different devices, we consider several devices with the same gate geometry, but with different SiGe spacer heights.
In the device with a 30 nm SiGe spacer, we choose an initial working point that corresponds to 1 electron in each quantum dot, and an interdot tunnel barrier of height 1 meV. We next apply the same CODA procedure used in Fig.~\ref{octo} to add one electron to the right dot, leaving the occupation of the left dot and the tunnel barrier height unchanged. 
We then increase the spacer height and repeat the procedure.  We note here that each new initial working point, corresponding to a new spacer height, was obtained from the previous initial point by allowing CODA to retune the appropriate electrode voltages, yielding a fully automated procedure.

The results of our CODA investigation, shown in Fig.~\ref{varyHeight}, help to compare and contrast the effectiveness of different device designs.  Since $L_1$ regularization yields solutions that are local and sparse, we expect that smaller values of $L_1$ will yield devices that are easier to control.  In Fig.~\ref{varyHeight}c, we see that reducing the height of the SiGe spacer layer improves the device controllability,
as measured by smaller magnitude changes in the control vectors (small $\Delta_V$).
This is intuitively reasonable because narrow spacers yield strong capacitive couplings between the dots and the top electrodes. In real devices, the advantages of narrow spacers may be offset by the proximity to disorder in the oxide, which is not included in our simulations.  However, it is clear that CODA can be an effective tool for assessing designs, and that many types of design questions can be addressed in this way.

In summary, we have introduced a protocol for the Compressed Optimization of Device Architectures, 
which applies an $L_1$ regularization scheme to a linearized model of device operation. 
We have demonstrated the effectiveness of this scheme by considering its application to semiconductor nanoelectronic quantum dot systems.
As devices continue to grow in complexity, such automated control schemes will be essential for design and operation.
Our protocol is computationally efficient to implement, 
and it provides a systematic approach for achieving local and sparse control.
Through realistic semiclassical simulations of double-dot devices, we have illustrated how the CODA scheme can be used for quantitative benchmarking and 
device development.
This method provides a path toward the rational design and operation of extensible quantum nanodevices.

\vspace{.2in}
\noindent \textbf{Methods}

We perform semi-classical Thomas-Fermi calculations\cite{Stopa:1996p13767} using the COMSOL Multiphysics software package to solve a nonlinear Poisson equation in three dimensions. We use zero-field boundary conditions on all sides of the simulated domain, with the exception of the bottom of the SiGe buffer, which is grounded.
We assume the following heterostructure profile for all the modeled devices, unless otherwise specified. This profile is consistent with the accumulation-mode devices described in refs. \onlinecite{Wu:2014p11938} and \onlinecite{Kawakami:2014p666}:
200~nm of Si$_{0.7}$Ge$_{0.3}$ (with dielectric constant $\varepsilon=13.19$),
a 10~nm Si quantum well ($\varepsilon=11.7$),
30~nm of Si$_{0.7}$Ge$_{0.3}$,
10~nm of Al$_2$O$_3$ ($\varepsilon=9.0$),
a 10~nm layer of metallic electrodes,
80~nm of Al$_2$O$_3$,
and a second 10~nm layer of metallic electrodes.

The occupations of the dots and the leads are found by integrating the induced 2D charge density over the accumulation regions. The heights of the tunnel barriers are calculated by numerically finding the saddle point 
over the appropriate potential energy landscape calculated by the self-consistent solver. By changing the voltage on the electrodes, the relevant occupations and tunnel barrier heights can be altered to achieve a desired working point, which in this case consists of one electron per dot and tunnel barrier heights of 1 meV. 
Small ($\sim$1 mV) voltage perturbations are then applied to each electrode, and the resulting changes in the operational targets are computed. These data are used to find the mapping in the equation $\hat S: \mathcal C \rightarrow \mathcal T$ of Eq.~\ref{eq1} (\emph{i.e.}, the cross-covariance matrix).

The $L_1$ and $L_2$ regularized optimization procedures are implemented using the CVXPY computational package\cite{JMLR:v17:15-408}.
For additional simulation details, see the Supplemental Information~\cite{suppMat}.

\emph{Acknowledgements--}
The authors thank J.~B.~Aidun, J.~Moussa, N.~T.~Jacobson, R.~P.~Muller, and C.~Tahan for useful comments and discussions.
This work was supported in part by the Laboratory Directed Research and Development program at Sandia National Laboratories and ARO (W911NF-12-1-0607 and W911NF-12-R-0012), NSF (PHY-1104660), ONR (N00014-15-1-0029).
Sandia National Laboratories is a multi-program laboratory managed and operated by Sandia Corporation, a wholly owned subsidiary of Lockheed Martin Corporation, for the U.S. Department of Energy's National Nuclear Security Administration under contract DE-AC04-94AL85000.
  
\emph{Author Contributions--}
JKG, AF, MF, SNC, RBK, and MAE developed the compressed control scheme. MAE, DRW, and JKG designed the devices. AF and JKG developed and carried out the device simulations. AF, JKG, MF, and SNC wrote the manuscript, with input from all authors.
 
\emph{Competing financial interests--} 
JKG, DRW, MAE, MF, and SNC are
co-inventors on a patent application related to some of the
nanostructure designs described in this work.
 
\emph{Additional Information--}
Supplementary information accompanies this paper. 
Correspondence and requests for materials should be addressed to John King Gamble (jkgambl\emph{@}sandia.gov).

\section{Supplemental Information}

\subsection{CODA procedure}
Here, we provide further details about the CODA protocol sketched in Fig.~1 of the main text. 
Suppose that we have $m$ operational targets and $n$ controls, and that $n>m$, so the system is underconstrained. 
We first identify a working point of experimental interest $(\mathbf c_\text{op}, \mathbf t_\text{op})$ and assume that perturbations of the device from the working point  are small.
We then linearize the mapping $\hat S: \mathcal C \rightarrow \mathcal T$ around $(\mathbf c_\text{op}, \mathbf t_\text{op})$; 
 we express the resulting linear map as a matrix $\hat S_\text{op}$.

In general, we define
\begin{equation}
\delta \mathbf c_\alpha = \underset{\delta \mathbf c} {\text{argmin}} \left|\left| \delta \mathbf c \right|\right|_\alpha,
\mbox{ subject to }\hat S_\text{op}(\mathbf c_\text{op} + \delta \mathbf c) = \mathbf t_\text{op}  + \delta \mathbf t \label{suppNorms},
\end{equation}
where 
$\delta \mathbf c$ ($\delta \mathbf t$) is the control (target) change from the working point, $|| \cdot ||_\alpha$ for $\alpha\geq1$ is the $L_n$ norm, and 
$|| \cdot ||_0$ is the $L_0$ pseudonorm\cite{Donoho:2006p1289}, which counts the number of non-zero elements in the vector. 
Although $\delta \mathbf c_0$ would be sparse by definition, finding $\delta \mathbf c_0$ is computationally infeasible for large systems, so we instead find $\delta \mathbf c_1$, which tends to be similarly sparse, as  shown in the field of compressed sensing\cite{Candes:2006p1207,Donoho:2006p1289}.

The full protocol for CODA is specified as follows:
\begin{enumerate}
\item Identify a working point $(\mathbf c_\text{op}, \mathbf t_\text{op})$ satisfying
$\hat S (\mathbf c_\text{op})=\mathbf t_\text{op}$.
In practice, $\hat S$ represents the action of some physical device, and $\mathbf t_\text{op}$ is the computed target vector corresponding to the input control $\mathbf c_\text{op}$ (e.g. voltages).

\item Consider a set of linearly independent, small control variations 
$\{\boldsymbol \epsilon_1,\boldsymbol \epsilon_2,...,\boldsymbol \epsilon_n\}$ ($\boldsymbol{\epsilon}_i \in \mathcal{C}$) about the working point.
(We took each $\boldsymbol \epsilon_j$ to correspond to a small voltage change on a single electrode.)
Perform $n$ simulations ${\hat S} (\mathbf c_\text{op} + \boldsymbol{\epsilon}_i)$ to obtain
the resulting $n$ data points $\mathbf t_\text{op}  + \delta \mathbf t_i$.
Since the control variations were small, by linearity ${\hat S}_\text{op} (\boldsymbol{\epsilon}_i)=\delta \mathbf t_i $.

\item
From the collection of $\{\boldsymbol \epsilon_1,\boldsymbol \epsilon_2,...,\boldsymbol \epsilon_n\}$ and the associated $\{\delta \mathbf t_1,\delta \mathbf t_2,...,\delta \mathbf t_n\}$, construct the linear operator ${\hat S}_\text{op} $ using least squares.

\item 
For a desired target variation $\delta \mathbf t$, identify the control variation $\delta \mathbf c_1$ as defined in Eq. \eqref{suppNorms}.
Using a convex program (such as the matrix-free cone solver implemented in the CVXPY package\cite{JMLR:v17:15-408}), find $\delta \mathbf c_1$ which has minimal $ \left| \left|  \delta \mathbf c \right| \right|_1$ subject to the constraint ${\hat S}_\text{op} (\delta \mathbf c)=\delta \mathbf t $.

\item Due to the nonlinearity of the system, in general $\hat S (\mathbf c_\text{op} + \delta \mathbf c_1) \neq \mathbf t_\text{op} + \delta \mathbf t$. However, because the linear approximation is valid within a neighborhood around the working point, there must exist a scalar $0<\gamma\leq 1$ such that $\hat S (\mathbf c_\text{op} + \gamma\delta \mathbf c_1) \approx S_\text{op} (\mathbf c_\text{op} + \gamma\delta \mathbf c_1)  = \mathbf t_\text{op} + \gamma\delta \mathbf t$. Use a series of simulations to find the maximum value of $\gamma$ for which this equation holds, $\gamma_0$. Define a new control working point $\mathbf c_\text{op}' =\mathbf c_\text{op} + \gamma_0\delta \mathbf c_1$, and new desired change in target $\delta \mathbf t' = (1-\gamma_0)\delta \mathbf t$. Repeat this procedure until convergence is achieved at the final target.
\end{enumerate}

\subsection{Simulation details}
We perform semi-classical Thomas-Fermi calculations\cite{Stopa:1996p13767} using the COMSOL Multiphysics software package to solve a nonlinear Poisson equation in three dimensions. Within the Si quantum well, we define a plane of charge with the charge density given by
\begin{equation}
\sigma_{\text{2D}}(x,y) = -2 \times 2 \times \frac{e m_{\textrm{eff}}(U(x,y)+E_F)}{2\pi \hbar^2} \times \theta(U(x,y)+E_F),
\end{equation}
where $e$ is the charge of an electron, $m_{\textrm{eff}} = 0.19\, m_{\text{electron}}$ is the transverse effective mass of a conduction electron in silicon, $U(x,y)$ is the strength of the electrostatic potential energy as a function of position, $E_F$ is the Fermi energy (which we take to be at ground), and $\theta(x)$ is the step function. The two prefactors account for the spin and valley degeneracies.

The details of the working point used in the analysis of the 8-dot device are given in the supplemental file \verb|8DotDevice.txt|. 
In this file, the physical attributes are listed first.
The dot occupations are expressed in electron numbers, and the tunnel barrier heights in eV. Voltages are given for each electrode, with the following labeling convention defined with respect to Fig.~2\textbf{a} of the main text.
Beginning with the upper layer of electrodes, Electrode~1 is in the lower-right corner of the schematic, and the ordering proceeds clockwise.
In the lower layer of gates, Electrode~5 is in the lower-right corner, and the ordering again proceeds clockwise.

Similar details for working points on the devices shown in Fig.~3 are given in the supplemental file \verb|VaryingHeight.txt|. As before, the dot occupations are given in numbers of electrons and the  tunnel barrier heights in eV. Here, the labeling convention for the electrodes begins with the upper layer at the electrode in the upper right corner of the schematic and proceeds clockwise.
On the lower layer of electrodes, the labeling begins at the electrode in the upper-left corner and proceeds clockwise.

Our CODA procedure requires all the components of the control vector to have the same units (and comparable magnitudes for numerical stability).
The operational targets considered in our simulations were electron occupations and tunnel barrier heights.
To make these quantities comparable for the devices studied here, we multiplied the tunnel barrier height by a factor of 10,  to ensure rapid convergence.

\bibliography{citations.bib}

\begin{thebibliography}{10}
\expandafter\ifx\csname url\endcsname\relax
  \def\url#1{\texttt{#1}}\fi
\expandafter\ifx\csname urlprefix\endcsname\relax\def\urlprefix{URL }\fi
\providecommand{\bibinfo}[2]{#2}
\providecommand{\eprint}[2][]{\url{#2}}

\bibitem{Petta:2005p2180}
\bibinfo{author}{Petta, J.~R.} \emph{et~al.}
\newblock \bibinfo{title}{Coherent manipulation of coupled electron spins in
  semiconductor quantum dots}.
\newblock \emph{\bibinfo{journal}{Science}} \textbf{\bibinfo{volume}{309}},
  \bibinfo{pages}{2180} (\bibinfo{year}{2005}).

\bibitem{Koppens:2005p717}
\bibinfo{author}{Koppens, F.} \emph{et~al.}
\newblock \bibinfo{title}{Control and detection of singlet-triplet mixing in a
  random nuclear field}.
\newblock \emph{\bibinfo{journal}{Science}} \textbf{\bibinfo{volume}{309}},
  \bibinfo{pages}{1346--1350} (\bibinfo{year}{2005}).

\bibitem{PhysRevLett.108.046808}
\bibinfo{author}{Prance, J.~R.} \emph{et~al.}
\newblock \bibinfo{title}{Single-shot measurement of triplet-singlet relaxation
  in a $\mathrm{Si}/\mathrm{SiGe}$ double quantum dot}.
\newblock \emph{\bibinfo{journal}{Phys. Rev. Lett.}}
  \textbf{\bibinfo{volume}{108}}, \bibinfo{pages}{046808}
  (\bibinfo{year}{2012}).

\bibitem{Maune:2012p344}
\bibinfo{author}{Maune, B.~M.} \emph{et~al.}
\newblock \bibinfo{title}{Coherent singlet-triplet oscillations in a
  silicon-based double quantum dot}.
\newblock \emph{\bibinfo{journal}{Nature}} \textbf{\bibinfo{volume}{481}},
  \bibinfo{pages}{344--347} (\bibinfo{year}{2012}).

\bibitem{Shulman:2012p202}
\bibinfo{author}{Shulman, M.~D.} \emph{et~al.}
\newblock \bibinfo{title}{Demonstration of entanglement of electrostatically
  coupled singlet-triplet qubits}.
\newblock \emph{\bibinfo{journal}{Science}} \textbf{\bibinfo{volume}{336}},
  \bibinfo{pages}{202} (\bibinfo{year}{2012}).

\bibitem{Pla:2012p541}
\bibinfo{author}{Pla, J.~J.} \emph{et~al.}
\newblock \bibinfo{title}{A single-atom electron spin qubit in silicon}.
\newblock \emph{\bibinfo{journal}{Nature}} \textbf{\bibinfo{volume}{489}},
  \bibinfo{pages}{541--545} (\bibinfo{year}{2012}).

\bibitem{Kim:2014p70}
\bibinfo{author}{Kim, D.} \emph{et~al.}
\newblock \bibinfo{title}{Quantum control and process tomography of a
  semiconductor quantum dot hybrid qubit}.
\newblock \emph{\bibinfo{journal}{Nature}} \textbf{\bibinfo{volume}{511}},
  \bibinfo{pages}{70--74} (\bibinfo{year}{2014}).

\bibitem{Wu:2014p11938}
\bibinfo{author}{Wu, X.} \emph{et~al.}
\newblock \bibinfo{title}{Two-axis control of a singlet--triplet qubit with an
  integrated micromagnet}.
\newblock \emph{\bibinfo{journal}{P. N. A. S.}} \textbf{\bibinfo{volume}{111}},
  \bibinfo{pages}{11938--11942} (\bibinfo{year}{2014}).

\bibitem{Fuechsle:2012p242}
\bibinfo{author}{Fuechsle, M.} \emph{et~al.}
\newblock \bibinfo{title}{A single-atom transistor}.
\newblock \emph{\bibinfo{journal}{Nature Nano.}} \textbf{\bibinfo{volume}{7}},
  \bibinfo{pages}{242--246} (\bibinfo{year}{2012}).

\bibitem{687879}
\bibinfo{author}{Koza, J.~R.}, \bibinfo{author}{Bennett, F.~H.},
  \bibinfo{author}{Andre, D.}, \bibinfo{author}{Keane, M.~A.} \&
  \bibinfo{author}{Dunlap, F.}
\newblock \bibinfo{title}{Automated synthesis of analog electrical circuits by
  means of genetic programming}.
\newblock \emph{\bibinfo{journal}{IEEE Transactions on Evolutionary
  Computation}} \textbf{\bibinfo{volume}{1}}, \bibinfo{pages}{109--128}
  (\bibinfo{year}{1997}).

\bibitem{Candes:2006p1207}
\bibinfo{author}{Cand{\`e}s, E.~J.}, \bibinfo{author}{Romberg, J.~K.} \&
  \bibinfo{author}{Tao, T.}
\newblock \bibinfo{title}{Stable signal recovery from incomplete and inaccurate
  measurements}.
\newblock \emph{\bibinfo{journal}{Communications on Pure and Applied
  Mathematics}} \textbf{\bibinfo{volume}{59}}, \bibinfo{pages}{1207--1223}
  (\bibinfo{year}{2006}).

\bibitem{Donoho:2006p1289}
\bibinfo{author}{Donoho, D.}
\newblock \bibinfo{title}{Compressed sensing}.
\newblock \emph{\bibinfo{journal}{IEEE Transactions on Information Theory}}
  \textbf{\bibinfo{volume}{52}}, \bibinfo{pages}{1289--1306}
  (\bibinfo{year}{2006}).

\bibitem{Hanson:2007p1217}
\bibinfo{author}{Hanson, R.}, \bibinfo{author}{Kouwenhoven, L.~P.},
  \bibinfo{author}{Petta, J.~R.}, \bibinfo{author}{Tarucha, S.} \&
  \bibinfo{author}{Vandersypen, L. M.~K.}
\newblock \bibinfo{title}{Spins in few-electron quantum dots}.
\newblock \emph{\bibinfo{journal}{Rev. Mod. Phys.}}
  \textbf{\bibinfo{volume}{79}}, \bibinfo{pages}{1217--1265}
  (\bibinfo{year}{2007}).

\bibitem{Loss:1998p120}
\bibinfo{author}{Loss, D.} \& \bibinfo{author}{DiVincenzo, D.~P.}
\newblock \bibinfo{title}{Quantum computation with quantum dots}.
\newblock \emph{\bibinfo{journal}{Phys. Rev. A}} \textbf{\bibinfo{volume}{57}},
  \bibinfo{pages}{120--126} (\bibinfo{year}{1998}).

\bibitem{Kane:1998p133}
\bibinfo{author}{Kane, B.~E.}
\newblock \bibinfo{title}{A silicon-based nuclear spin quantum computer}.
\newblock \emph{\bibinfo{journal}{Nature}} \textbf{\bibinfo{volume}{393}},
  \bibinfo{pages}{133--137} (\bibinfo{year}{1998}).

\bibitem{Medford:2013p654}
\bibinfo{author}{Medford, J.} \emph{et~al.}
\newblock \bibinfo{title}{Self-consistent measurement and state tomography of
  an exchange-only spin qubit}.
\newblock \emph{\bibinfo{journal}{Nature Nano.}} \textbf{\bibinfo{volume}{8}},
  \bibinfo{pages}{654--659} (\bibinfo{year}{2013}).

\bibitem{Medford:2013p050501}
\bibinfo{author}{Medford, J.} \emph{et~al.}
\newblock \bibinfo{title}{Quantum-dot-based resonant exchange qubit}.
\newblock \emph{\bibinfo{journal}{Phys. Rev. Lett.}}
  \textbf{\bibinfo{volume}{111}}, \bibinfo{pages}{050501}
  (\bibinfo{year}{2013}).

\bibitem{Kawakami:2014p666}
\bibinfo{author}{Kawakami, E.} \emph{et~al.}
\newblock \bibinfo{title}{Electrical control of a long-lived spin qubit in a
  si/sige quantum dot}.
\newblock \emph{\bibinfo{journal}{Nature Nano.}} \textbf{\bibinfo{volume}{9}},
  \bibinfo{pages}{666--670} (\bibinfo{year}{2014}).

\bibitem{Veldhorst:preprint2013}
\bibinfo{author}{Veldhorst, M.} \emph{et~al.}
\newblock \bibinfo{title}{An addressable quantum dot qubit with fault-tolerant
  control fidelity}.
\newblock \emph{\bibinfo{journal}{Nature Nano.}} \textbf{\bibinfo{volume}{9}},
  \bibinfo{pages}{981--985} (\bibinfo{year}{2014}).

\bibitem{Enge1500214}
\bibinfo{author}{Eng, K.} \emph{et~al.}
\newblock \bibinfo{title}{Isotopically enhanced triple-quantum-dot qubit}.
\newblock \emph{\bibinfo{journal}{Science Advances}}
  \textbf{\bibinfo{volume}{1}}, \bibinfo{pages}{e1500214}
  (\bibinfo{year}{2015}).

\bibitem{KimNatNano15}
\bibinfo{author}{Kim, D.} \emph{et~al.}
\newblock \bibinfo{title}{Microwave-driven coherent operation of a
  semiconductor quantum dot charge qubit}.
\newblock \emph{\bibinfo{journal}{Nature Nano.}} \textbf{\bibinfo{volume}{10}},
  \bibinfo{pages}{243--247} (\bibinfo{year}{2015}).

\bibitem{Kim:2015aa}
\bibinfo{author}{Kim, D.} \emph{et~al.}
\newblock \bibinfo{title}{High-fidelity resonant gating of a silicon-based
  quantum dot hybrid qubit}.
\newblock \emph{\bibinfo{journal}{Npj Quantum Information}}
  \textbf{\bibinfo{volume}{1}}, \bibinfo{pages}{15004} (\bibinfo{year}{2015}).

\bibitem{PhysRevLett.116.046802}
\bibinfo{author}{Delbecq, M.~R.} \emph{et~al.}
\newblock \bibinfo{title}{Quantum dephasing in a gated gaas triple quantum dot
  due to nonergodic noise}.
\newblock \emph{\bibinfo{journal}{Phys. Rev. Lett.}}
  \textbf{\bibinfo{volume}{116}}, \bibinfo{pages}{046802}
  (\bibinfo{year}{2016}).

\bibitem{2016arXiv160207833T}
\bibinfo{author}{{Takeda}, K.} \emph{et~al.}
\newblock \bibinfo{title}{{A fault-tolerant addressable spin qubit in a natural
  silicon quantum dot}} (\bibinfo{year}{2016}).
\newblock \bibinfo{note}{Preprint at http://arxiv.org/abs/1602.07833}.

\bibitem{PhysRevLett.116.116801}
\bibinfo{author}{Martins, F.} \emph{et~al.}
\newblock \bibinfo{title}{Noise suppression using symmetric exchange gates in
  spin qubits}.
\newblock \emph{\bibinfo{journal}{Phys. Rev. Lett.}}
  \textbf{\bibinfo{volume}{116}}, \bibinfo{pages}{116801}
  (\bibinfo{year}{2016}).

\bibitem{Pla:2013p334}
\bibinfo{author}{Pla, J.~J.} \emph{et~al.}
\newblock \bibinfo{title}{High-fidelity readout and control of a nuclear spin
  qubit in silicon}.
\newblock \emph{\bibinfo{journal}{Nature}} \textbf{\bibinfo{volume}{496}},
  \bibinfo{pages}{334--338} (\bibinfo{year}{2013}).

\bibitem{Muhonen:2014p1402.7140}
\bibinfo{author}{Muhonen, J.~T.} \emph{et~al.}
\newblock \bibinfo{title}{Storing quantum information for 30 seconds in a
  nanoelectronic device}.
\newblock \emph{\bibinfo{journal}{Nature Nano.}} \textbf{\bibinfo{volume}{9}},
  \bibinfo{pages}{986--991} (\bibinfo{year}{2014}).

\bibitem{Zwanenburg:2013p961}
\bibinfo{author}{Zwanenburg, F.~A.} \emph{et~al.}
\newblock \bibinfo{title}{Silicon quantum electronics}.
\newblock \emph{\bibinfo{journal}{Rev. Mod. Phys.}}
  \textbf{\bibinfo{volume}{85}}, \bibinfo{pages}{961} (\bibinfo{year}{2013}).

\bibitem{2015arXiv151201606H}
\bibinfo{author}{{Harvey-Collard}, P.} \emph{et~al.}
\newblock \bibinfo{title}{{Nuclear-driven electron spin rotations in a single
  donor coupled to a silicon quantum dot}} (\bibinfo{year}{2015}).
\newblock \bibinfo{note}{Preprint at http://arxiv.org/abs/1512.01606}.

\bibitem{0953-8984-27-15-154205}
\bibinfo{author}{Muhonen, J.~T.} \emph{et~al.}
\newblock \bibinfo{title}{Quantifying the quantum gate fidelity of single-atom
  spin qubits in silicon by randomized benchmarking}.
\newblock \emph{\bibinfo{journal}{Journal of Physics: Condensed Matter}}
  \textbf{\bibinfo{volume}{27}}, \bibinfo{pages}{154205}
  (\bibinfo{year}{2015}).

\bibitem{Lauchte1500022}
\bibinfo{author}{Laucht, A.} \emph{et~al.}
\newblock \bibinfo{title}{Electrically controlling single-spin qubits in a
  continuous microwave field}.
\newblock \emph{\bibinfo{journal}{Science Advances}}
  \textbf{\bibinfo{volume}{1}}, \bibinfo{pages}{e1500022}
  (\bibinfo{year}{2015}).

\bibitem{PhysRevLett.112.240504}
\bibinfo{author}{Kelly, J.} \emph{et~al.}
\newblock \bibinfo{title}{Optimal quantum control using randomized
  benchmarking}.
\newblock \emph{\bibinfo{journal}{Phys. Rev. Lett.}}
  \textbf{\bibinfo{volume}{112}}, \bibinfo{pages}{240504}
  (\bibinfo{year}{2014}).

\bibitem{Schulman:2014unpub}
\bibinfo{author}{Shulman, M.~D.} \emph{et~al.}
\newblock \bibinfo{title}{Suppressing qubit dephasing using real-time
  {H}amiltonian estimation}.
\newblock \emph{\bibinfo{journal}{Nat. Commun.}} \textbf{\bibinfo{volume}{5}},
  \bibinfo{pages}{5156} (\bibinfo{year}{2014}).

\bibitem{BaartAPL}
\bibinfo{author}{Baart, T.~A.}, \bibinfo{author}{Eendebak, P.~T.},
  \bibinfo{author}{Reichl, C.}, \bibinfo{author}{Wegscheider, W.} \&
  \bibinfo{author}{Vandersypen, L. M.~K.}
\newblock \bibinfo{title}{Computer-automated tuning of semiconductor double
  quantum dots into the single-electron regime}.
\newblock \emph{\bibinfo{journal}{Applied Physics Letters}}
  \textbf{\bibinfo{volume}{108}}, \bibinfo{pages}{213104}
  (\bibinfo{year}{2016}).

\bibitem{Natarajan:1995p227}
\bibinfo{author}{Natarajan, B.~K.}
\newblock \bibinfo{title}{Sparse approximate solutions to linear systems}.
\newblock \emph{\bibinfo{journal}{SIAM Journal on Computing}}
  \textbf{\bibinfo{volume}{24}}, \bibinfo{pages}{227--234}
  (\bibinfo{year}{1995}).

\bibitem{suppMat}
\bibinfo{author}{Frees, A.} \emph{et~al.}
\newblock \bibinfo{title}{Supplemental information}.
\newblock \emph{\bibinfo{journal}{Supplemental Information}}
  (\bibinfo{year}{2016}).

\bibitem{ward:2016}
\bibinfo{author}{Ward, D.} \emph{et~al.}
\newblock \bibinfo{title}{State-conditional coherent charge qubit oscillations
  in a si/sige quadruple quantum dot}.
\newblock \emph{\bibinfo{journal}{arXiv preprint arXiv:1604.07956}}
  (\bibinfo{year}{2016}).

\bibitem{Stopa:1996p13767}
\bibinfo{author}{Stopa, M.}
\newblock \bibinfo{title}{Quantum dot self-consistent electronic structure and
  the coulomb blockade}.
\newblock \emph{\bibinfo{journal}{Phys. Rev. B}} \textbf{\bibinfo{volume}{54}},
  \bibinfo{pages}{13767--13783} (\bibinfo{year}{1996}).

\bibitem{2015Nanot..26t5703S}
\bibinfo{author}{{Shirkhorshidian}, A.} \emph{et~al.}
\newblock \bibinfo{title}{{Transport spectroscopy of low disorder silicon
  tunnel barriers with and without Sb implants}}.
\newblock \emph{\bibinfo{journal}{Nanotechnology}}
  \textbf{\bibinfo{volume}{26}}, \bibinfo{pages}{205703}
  (\bibinfo{year}{2015}).

\bibitem{JMLR:v17:15-408}
\bibinfo{author}{Diamond, S.} \& \bibinfo{author}{Boyd, S.}
\newblock \bibinfo{title}{Cvxpy: A python-embedded modeling language for convex
  optimization}.
\newblock \emph{\bibinfo{journal}{Journal of Machine Learning Research}}
  \textbf{\bibinfo{volume}{17}}, \bibinfo{pages}{1--5} (\bibinfo{year}{2016}).

\end{thebibliography}

\end{document}